\documentclass[aps,prx,reprint,superscriptaddress]{revtex4-2}

\usepackage{amsfonts}
\usepackage{amssymb}
\usepackage{physics}
\usepackage{hyperref}
\usepackage{graphicx}

\begin{document}

\title{A Unified Categorical Description of Quantum Hall Hierarchy and Anyon Superconductivity}

\author{Donghae Seo}
\affiliation{Department of Physics, Pohang University of Science and Technology, Pohang 37673, Republic of Korea}

\author{Taegon Lee}
\affiliation{Department of Physics, Korea Advanced Institute of Science and Technology, Daejeon 34141, Republic of Korea}
 
\author{Gil Young Cho}
\email{gilyoungcho@kaist.ac.kr}
\affiliation{Department of Physics, Korea Advanced Institute of Science and Technology, Daejeon 34141, Republic of Korea}
\affiliation{Center for Artificial Low Dimensional Electronic Systems, Institute for Basic Science, Pohang 37673, Republic of Korea}
 
\date{\today}

\begin{abstract}
We present a unified category-theoretic framework for quantum Hall hierarchy constructions and anyon superconductivity based on modular tensor categories over $\mathrm{Rep}(\mathrm{U}(1))$ and $\mathrm{sRep}(\mathrm{U}(1)^f)$. Our approach explicitly incorporates conserved $\mathrm{U}(1)$ charge and formulates doping via a generalized stack-and-condense procedure, in which an auxiliary topological order is stacked onto the parent phase, and the quasiparticles created by doping subsequently condense. Depending on whether this condensation preserves or breaks the $\mathrm{U}(1)$ symmetry, the system undergoes a transition to a quantum Hall hierarchy state or to an anyon superconductor. For anyon superconductors, the condensate charge is determined unambiguously by the charged local bosons contained in the condensable algebra. Our framework reproduces all known anyon superconductors obtained from field-theoretic analyses and further predicts novel phases, including a charge-$2e$ anyon superconductor derived from the Laughlin state and charge-$ke$ anyon superconductors arising from bosonic $\mathbb{Z}_k$ Read-Rezayi states. By placing hierarchy transitions and anyon superconductivity within a single mathematical formalism, our work provides a unified understanding of competing and proximate phases near experimentally realizable fractional quantum Hall states.
\end{abstract}

\maketitle

\section{Introduction} 

Anyons are emergent quasiparticles in two-dimensional systems whose exchange statistics are neither bosonic nor fermionic \cite{wen2004quantum}. They arise in strongly correlated phases such as fractional quantum Hall states under strong magnetic fields and in some quantum spin liquids of geometrically frustrated magnets \cite{wen2017colloquium}. Recently, highly tunable moiré platforms, namely twisted bilayer MoTe$_2$ and multilayer rhombohedral graphene, have emerged as new settings for anyonic physics \cite{cai2023signatures,zeng2023thermodynamic,park2023observation,xu2023observation,ji2024local,redekop2024direct,reddy2024non-abelian,ahn2024non-Abelian,lu2024fractional,kang2024evidence,lu2025extended,chen2025robust,xu2025multiple,wang2025higher,liu2025non-abelian,liu2025parafermions,xie2025tunable,aronson2025dispacement,kang2025time-reversal,park2025observation}. These systems realize fractional Chern insulators, which are lattice analogues of fractional quantum Hall states in the absence of external magnetic fields. There, anyonic excitations can acquire finite dispersion \cite{shi2024doping,pichler2025microscopic,schleith2025anyon,yan2025anyon} and, under electrostatic gating, appear at finite density, enabling collective phenomena driven by interacting anyons beyond the conventional paradigms of bosons or fermions. A particularly intriguing possibility is superconductivity induced by doping anyons, known as anyon superconductivity \cite{laughlin1988relationship,laughlin1988superconducting,fetter1989random-phase,wen1989chiral,chen1989anyon,lee1989anyon,hosotani1990superconductivity,wen1990compressibility,lee1991anyon,shi2024doping,pichler2025microscopic,divic2025anyon,shi2025doping,han2025anyon,zhang2025charge,kuhlenkamp2025robust,shi2025nonabelian,chen2025finite-momentum,wang2025chiral,zhang2025holon,ahn2025superconductivity,yoon2026quarter-metal}. This scenario has attracted significant recent interest \cite{shi2024doping,pichler2025microscopic,divic2025anyon,shi2025doping,han2025anyon,zhang2025charge,kuhlenkamp2025robust,shi2025nonabelian,chen2025finite-momentum,wang2025chiral,zhang2025holon,ahn2025superconductivity,yoon2026quarter-metal}, motivated in part by experimental observations of superconductivity in these moiré materials \cite{han2025signatures,xu2025signatures}. While this exotic superconductivity closely resembles the fractional quantum Hall hierarchy, where new topological phases arise from the formation of additional quantum Hall fluids of anyonic excitations on top of a parent fractional quantum Hall state \cite{haldane1983fractional,halperin1984statistics,hansson2017quantum}, the precise relationship between the two anyon-driven phenomena has yet to be fully clarified.

\begin{figure*}
    \centering
    \includegraphics[width=0.9\textwidth]{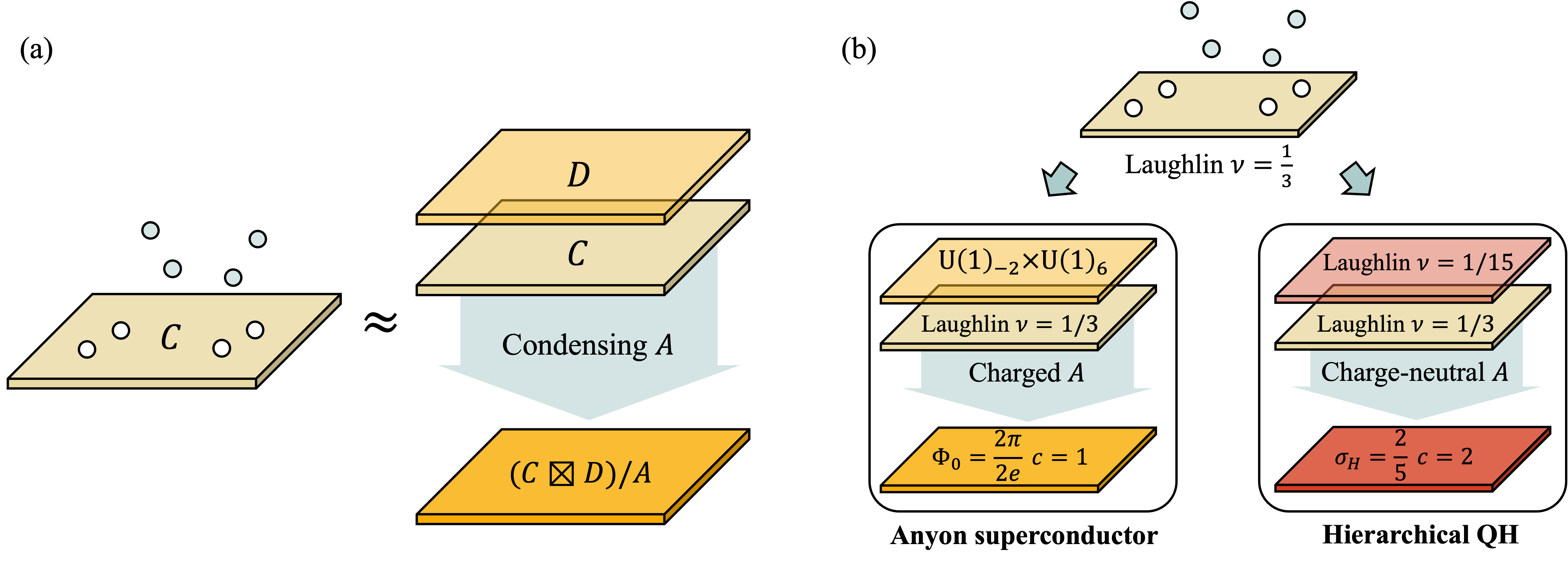}
    \caption{(a) Schematic description of the stack-and-condense procedure. When a given $\mathrm{U}(1)$-symmetric topological order $\mathcal{C}$ is doped, the emerging anyons form their own topological state $\mathcal{D}$, which is stacked onto $\mathcal{C}$, and then a condensable algebra $\mathcal{A}$ including the doped anyon condenses. The final state is described by $(\mathcal{C} \boxtimes \mathcal{D}) / \mathcal{A}$. (b) Anyon superconductor and hierarchical quantum Hall (QH) state from doping the Laughlin state at filling fraction $\nu = \frac{1}{3}$. The Haldane-Halperin hierarchical state with filling fraction $\nu = \frac{2}{5}$ and chiral central charge $c = 2$ is obtained by choosing $\mathcal{D}$ as the Laughlin state at filling fraction $\nu = \frac{1}{15}$ and condensing a charge-neutral $\mathcal{A}$. A chiral charge-$2e$ anyon superconductor with $c = 1$ is obtained by choosing $\mathcal{D}$ as the charge-neutral $\mathrm{U}(1)_{-2} \times \mathrm{U}(1)_6$ theory and condensing a charged $\mathcal{A}$ including the charge-$2$ local boson $\mathbf{2}$. In the figure, $\Phi_0$ denotes the magnetic flux quantum of the superconductor.}
    \label{fig:schematic}
\end{figure*}

In this manuscript, we introduce a categorical description of anyon superconductivity that places fractional quantum Hall hierarchy states and anyon superconductivity on equal theoretical footing. This perspective is motivated by the shared feature that both phenomena arise from anyonic low-energy degrees of freedom, whose structure is governed by a modular tensor category \cite{wen2016theory,lan2016theory,lan2017classification,bruillard2017fermionic,cho2023classification,ng2025classification}. To describe anyon superconductivity, it is essential to track electric charge and to generalize conventional modular tensor categories with finitely many simple objects to incorporate the global $\mathrm{U}(1)$ charge conservation. This leads to modular tensor categories over either $\mathrm{Rep}(\mathrm{U}(1))$ or $\mathrm{sRep}(\mathrm{U}(1)^f)$, which contain infinitely many simple objects. Within this framework, transitions to both fractional quantum Hall hierarchy states and anyon superconductors are naturally described by a stack-and-condense operation \cite{zhang2025hierarchy,shi2025doping}. Our protocol is schematically illustrated in Fig.~\ref{fig:schematic}.

Our central contribution is an extension of the stack-and-condense operation to incorporate global $\mathrm{U}(1)$ charge conservation, yielding a direct and systematic method for determining the electric charge of the condensate in an anyon superconductor. This generalization builds upon the stack-and-condense framework originally developed for quantum Hall hierarchy constructions \cite{zhang2025hierarchy} and subsequently applied to certain cases of anyon superconductivity within conventional modular tensor category settings \cite{shi2025doping}. In an anyon superconductor, the condensate charge generally cannot be inferred directly from the charge of the doped anyons, as the two often differ \cite{shi2024doping,shi2025doping}; for example, doping charge-$e/4$ anyons in a Pfaffian state yields a conventional charge-$2e$ superconductor \cite{shi2025doping}. Within our categorical formulation, the condensate charge of the superconducting phase is uniquely fixed by the symmetry-breaking pattern of the global $\mathrm{U}(1)$ symmetry, and the chiral central charge follows directly within the same framework.

Notably, we find that our framework reproduces known field-theoretical results on anyon superconductors, including those derived from non-Abelian parent states \cite{shi2025doping}, while also predicting new classes of anyonic phases. These include charge-$2e$ anyon superconductivity emerging from the Laughlin state at filling fraction $\nu = \frac{1}{3}$ and a variety of charge-$ke$ anyon superconductors derived from bosonic $\mathbb{Z}_k$ Read-Rezayi states. Taken together, these results establish a unified framework for anyon superconductivity and hierarchy constructions, providing a systematic bridge between categorical data, field-theoretic descriptions, and experimentally relevant anyonic phases.

The remainder of this paper is organized as follows. In Sec.~\ref{sec:categorical}, we motivate our approach by presenting a field-theoretic analysis of the Laughlin state at filling fraction $\nu = \frac{1}{3}$ that integrates the hierarchy construction and anyon superconductivity. We then establish the formal categorical framework used throughout this work. Sections~\ref{sec:fermionic} and \ref{sec:bosonic} provide explicit examples of various anyon superconductors formulated within our categorical framework for fermionic and bosonic systems, respectively. We summarize our results and conclude in Sec.~\ref{sec:conclusion}. For the completeness of the manuscript, the $K$-matrix formulation \cite{wen1992classification} of Abelian topological orders is briefly reviewed in App.~\ref{app:K-matrix} and field-theoretic analyses of anyon superconductors \cite{shi2024doping,shi2025doping} are provided in App.~\ref{app:review}. Additional mathematical details, including background on modular tensor category, symmetry enrichment, and anyon condensation, are provided in Apps.~\ref{app:mtc}, \ref{app:categorical}, and \ref{app:anyon}.

\section{Categorical Framework} \label{sec:categorical}

\subsection{Laughlin state as a motivating example} \label{sec:Laughlin}

We begin with a field theory example in which both superconducting and hierarchy states emerge from a single parent topological order, namely the Laughlin state at the filling fraction $\nu = \frac{1}{3}$, thereby motivating a unified categorical approach.

The Laughlin state is described by \cite{wen2004quantum}
\begin{align}
    \mathcal{L}_\mathrm{Laughlin} &= - \frac{3}{4 \pi} \epsilon^{\mu\nu\lambda} a_\mu \partial_\nu  a_\lambda  + \frac{1}{2 \pi} \epsilon^{\mu\nu\lambda} A_\mu \partial_\nu a_\lambda, \nonumber \\ 
    & \equiv - \frac{3}{4 \pi} ada  + \frac{1}{2 \pi} Ada. 
\end{align}
where the repeated indices in the first line are summed over. Here $a_\mu$ is a dynamical $\mathrm{U}(1)$ gauge field, $A_\mu$ is the background electromagnetic gauge potential. From the first to the second line, we use standard differential-form notation, with gauge fields as one-forms and wedge products between them. The anyons can be incorporated by introducing a conserved current $j$ that minimally couples to $a$, $j = \frac{1}{2\pi} d\alpha$, with $\alpha$ an additional $\mathrm{U}(1)$ gauge field.

In the standard hierarchy construction \cite{wen2004quantum}, we endow a level-$m$ Chern-Simons term to $\alpha$: 
\begin{align} \label{eq:Laughlin_hierarchy}
    \mathcal{L}_\mathrm{hierarchy} &= - \frac{3}{4 \pi} a d a + \frac{1}{2 \pi} A d a  + \frac{1}{2 \pi} a d \alpha - \frac{m}{4 \pi} \alpha d \alpha.
\end{align}
Setting $m=-2$ realizes the familiar $\nu=\frac{2}{5}$ hierarchy state. In the $K$-matrix formalism \cite{wen1992classification}, Eq.~\eqref{eq:Laughlin_hierarchy} is characterized by 
\begin{align} \label{eq:Kmatrix}
    K = 
    \begin{pmatrix}
        3 & -1 \\
        -1 & m
    \end{pmatrix}, \quad 
    \mathbf{t} = 
    \begin{pmatrix}
        1 \\ 0
    \end{pmatrix}
\end{align}
where $\mathbf{t}$ is the charge vector (see Appendix~\ref{app:K-matrix} for details). 

Within the same hierarchical construction, a superconducting state can also be realized. This follows from the observation that setting $m=\frac{1}{3}$ renders the $K$ matrix in Eq.~\eqref{eq:Kmatrix} singular, with zero determinant, signaling the emergence of a superconductor. However, Chern-Simons levels cannot be fractional. Thus, $m = \frac{1}{3}$ should be viewed as an effective response theory obtained by integrating out some dynamical gauge fields. The desired effective level $m = \frac{1}{3}$ can be realized by the $\mathrm{U}(1)_{-2} \times \mathrm{U}(1)_6$ Chern-Simons theory: 
\begin{align}
    \mathcal{L}_{\mathrm{U}(1)_{-2} \times \mathrm{U}(1)_6} &= \frac{2}{4 \pi} b d b + \frac{1}{2 \pi} \alpha d b - \frac{6}{4 \pi} c d c + \frac{1}{2 \pi} \alpha d c, 
\end{align}
where $b$ and $c$ are dynamical $\mathrm{U}(1)$ gauge fields. The full Lagrangian is characterized by 
\begin{align} \label{eq:Laughlin_anyon_superconductor_K}
    K = 
    \begin{pmatrix}
        3 & -1 & 0 & 0 \\
        -1 & 0 & -1 & -1 \\
        0 & -1 & -2 & 0 \\
        0 & -1 & 0 & 6
    \end{pmatrix}, \quad 
    \mathbf{t} = 
    \begin{pmatrix}
        1 \\ 0 \\ 0 \\ 0
    \end{pmatrix}.
\end{align}
The $K$-matrix has zero determinant as desired. Performing an $\mathrm{SL}(4,\mathbb{Z})$ basis transformation and integrating out the massive gauge fields associated with the trivial topological sector (see Appendix~\ref{app:details}), we obtain
\begin{align} \label{eq:Laughlin_anyon_superconductor}
\mathcal{L}_{\mathrm{SC}} = \frac{2}{2\pi} A d\tilde{a} + \frac{1}{4\pi} A dA + 2\mathrm{CS}_g,
\end{align}
where $\mathrm{CS}_g$ denotes the gravitational Chern-Simons term. This describes a charge-$2e$ superconductor with chiral central charge $c = 1$, realized using the same framework as the hierarchy construction. This example demonstrates the correspondence between anyon superconductivity and quantum Hall hierarchy constructions, which will be elucidated by the categorical approach.

\subsection{Modular tensor category over \texorpdfstring{$\mathrm{sRep}(\mathrm{U}(1)^f)$}{}}

We now introduce modular tensor categories over $\mathrm{sRep}(\mathrm{U}(1)^f)$—that is, fermionic modular tensor categories enriched by global $\mathrm{U}(1)$ charge conservation—as the natural mathematical framework for describing {both the} anyon superconductivity {and hierarchy construction} in electronic systems. For bosonic systems, the corresponding framework is given by modular tensor categories over $\mathrm{Rep}(\mathrm{U}(1))$. While the discussion that follows is necessarily formal, we present it in a pedagogical manner to maintain clarity and accessibility. A brief introduction to the conventional modular tensor categories is provided in App.~\ref{app:mtc}.

The two-dimensional topological orders with an onsite finite symmetry $G$ are characterized by modular tensor categories $\mathcal{C}$ over the symmetric fusion category $\mathcal{E}=\mathrm{Rep}(G)$ for bosonic systems or $\mathrm{sRep}(G^f)$ for fermionic systems~\cite{lan2017classification,lan2017modular}. Here we extend this formulation to a continuous symmetry, namely the $\mathrm{U}(1)$ symmetry associated with global charge conservation. Since $\mathrm{U}(1)$ has infinitely many irreducible representations labeled by integers—corresponding to electric charge in our context—the symmetry fusion category $\mathcal{E} = \mathrm{Rep}(\mathrm{U}(1))$ (or $\mathrm{sRep}(\mathrm{U}(1)^f)$)
contains infinitely many simple objects, which we denote by integers in bold font 
\begin{align}
\cdots, \mathbf{\bar{2}}, \mathbf{\bar{1}}, \mathbf{0}, \mathbf{1}, \mathbf{2}, \cdots, 
\end{align}
where the overbar denotes a minus sign. Physically, a simple object $\mathbf{m}$ corresponds to a local excitation carrying charge $m$, in units of fundamental electron charge $e$. They braid trivially with one another and obey the Abelian fusion rule given by the ordinary addition of integers, e.g., $\mathbf{2} \otimes \mathbf{\bar{1}} = \mathbf{1}$. In systems that obey the spin-charge relation \cite{seiberg2016gapped}, as is typical in condensed matter, excitations with odd electric charge must exhibit fermionic self-statistics. Conversely, excitations with even electric charge exhibit bosonic self-statistics. In bosonic systems, all local particles $\mathbf{m}$ with integer charges {are} bosonic. Some technical details of the symmetric fusion category are delegated to Appendix~\ref{app:categorical}.

The anyons in a $\mathrm{U}(1)$-symmetric topological order can carry fractional charges. In conventional categorical formalism \cite{barkeshli2019symmetry,cheng2016translational}, the fractional charges are determined modulo $1$ via mutual statistics phase with the vison \cite{barkeshli2019symmetry,cheng2016translational}. In the modular tensor categories over $\mathrm{Rep}(\mathrm{U}(1))$ or $\mathrm{sRep}(\mathrm{U}(1)^f)$, {by} contrast, the charges of anyons are determined by their fusion rules. Let us consider an anyon $a$ in a given $\mathrm{U}(1)$-symmetric topological order. By repeated fusion of $a$, a local excitation can be generated. The smallest number of fusions under which we first get a local excitation is called the order $n_a$ of $a$, such that
\begin{align}
    a^{n_a} = k_\mathbf{m} \mathbf{m} \oplus \cdots,  
\end{align}
where $k_\mathbf{m}$ is a positive integer coefficient. Since $\mathbf{m}$ carries charge $m$ and the charges are conserved under fusion, we can determine that the charge of $a$ as $Q_a = \frac{m}{n_a}$ without taking it modulo one. When considered modulo $1$, this charge assignment is consistent with the charge inferred from the mutual statistical phase with the vison. 

For a pedagogical example, we can consider the Laughlin state at filling $\nu = \frac{1}{3}$. The Laughlin state is known to host Abelian anyons $a_{\frac{1}{3}}$ and $a_{\frac{2}{3}}$, where the subscripts denote the charges of the anyons. Since the Laughlin state obeys the spin-charge relation, its local excitations are described by $\mathrm{sRep}(\mathrm{U}(1)^f) = \{\cdots, \Bar{\mathbf{1}}, \mathbf{0}, \mathbf{1}, \mathbf{2}, \cdots\}$ where the odd-integer objects like $\Bar{\mathbf{1}}$ or $\mathbf{1}$ are fermions. The anyons satisfy 
\begin{align}
a_{\frac{1}{3}} \otimes a_{\frac{1}{3}} = a_{\frac{2}{3}}, \quad a_{\frac{1}{3}} \otimes a_{\frac{2}{3}} = \mathbf{1}.  
\end{align} 
The anti-anyon of $a_{\frac{1}{3}}$, denoted by $a_{\frac{1}{3}}^{-1}$, carries charge $-\frac{1}{3}$ and satisfies 
\begin{align}
    a_{\frac{1}{3}} \otimes a_{\frac{1}{3}}^{-1} = \mathbf{0}. 
\end{align}
We may also consider the fusion of $a_{\frac{1}{3}}$ and an electron $\mathbf{1}$  
\begin{align}
    a_{\frac{1}{3}} \otimes \mathbf{1} = \mathbf{1} a_{\frac{1}{3}}, 
\end{align}
where $\mathbf{1} a_{\frac{1}{3}}$ intuitively denotes the anyon carrying electric charge $\frac{4}{3} = 1 + \frac{1}{3}$. 

We note in passing that modular tensor categories over $\mathrm{Rep}(\mathrm{U}(1))$ or $\mathrm{sRep}(\mathrm{U}(1)^f)$ in fact have appeared in literature in different contexts. In Ref.~\cite{cheng2022gauging}, the infinite category, which is denoted by $\mathcal{C}'$ by the authors, appears as an intermediate step towards gauging $\mathrm{U}(1)$ symmetry. In Ref.~\cite{lan2017hierarchy}, a similar mathematical structure appears during computing non-Abelian families of topological orders. In these references~\cite{cheng2022gauging,lan2017hierarchy}, infinite categories appear only as an intermediate or virtual structure during the calculation. In contrast, we identify the objects of the category with physical, particle-like excitations, which will be proven crucial for anyon superconductivity.

\subsection{Anyon condensation with \texorpdfstring{$\mathrm{U}(1)$}{} charge}

Like ordinary condensation of bosonic particles, anyons with bosonic self-statistics can condense to form a new ground state. Mathematically, such condensation processes are classified by condensable algebra \cite{kong2014anyon} in the corresponding modular tensor category (see Appendix~\ref{app:anyon}). 

Anyons that do not possess bosonic self-statistics cannot condense on their own. To describe transitions driven by the condensation of such anyons, the stack-and-condense construction was introduced in Refs.~\cite{zhang2025hierarchy,shi2025doping}. In this approach, an additional layer of a topological order $\mathcal{D}$ is first stacked onto the parent phase $\mathcal{C}$, allowing an anyon from the stacked layer to bind with an anyon from the parent layer to form a condensable boson (Fig.~\ref{fig:schematic}). In the case of Abelian topological orders, this stack-and-condense construction can be justified using a field-theoretic parton approach~\cite{shi2025doping}, while remaining applicable to non-Abelian orders as well. Moreover, when considering systems composed of electrons, both $\mathcal{C}$ and $\mathcal{D}$ must obey the spin-charge relation. This requirement immediately implies that, if the stacked theory $\mathcal{D}$ is charge-neutral, i.e., all anyons in $\mathcal{D}$ are charge-neutral, $\mathcal{D}$ must be a bosonic topological order. Otherwise, the combined system would contain a charge-neutral local fermion, thereby violating the spin-charge relation.

More specifically, suppose we wish to condense a general anyon $a$ in a given topological order $\mathcal{C}$. To achieve this, we stack an additional topological order $\mathcal{D}$ containing an anyon $b$ such that the composite $ab$ forms a condensable boson in the product theory $\mathcal{C} \boxtimes \mathcal{D}$. We then condense the composite anyon $ab$ to obtain the final theory, which is described by $\left(\mathcal{C} \boxtimes \mathcal{D}\right) / \mathcal{A}$, where $\mathcal{A}$ is the condensable algebra ``generated'' by $ab$. Specifically, every anyon $c \in \mathcal{A}$ appears in some repeated fusion of $ab$; that is, for each $c \in \mathcal{A}$ there exists an integer $k$ such that $(ab)^k = c \oplus \cdots$. The computation of $\left(\mathcal{C}\boxtimes\mathcal{D}\right)/\mathcal{A}$ follows the standard theory of anyon condensation \cite{kong2014anyon} in conventional modular tensor categories (see Appendix~\ref{app:anyon}). 

We now illustrate how this approach unifies anyon superconductivity and the quantum Hall hierarchy construction. As an example, we consider a fermionic topological order $\mathcal{C}$ described by a modular tensor category over $\mathrm{sRep}(\mathrm{U}(1)^f)$. Since we focus on descendant phases of the fermionic $\mathrm{U}(1)$-symmetric topological orders, the product theory $\mathcal{C} \boxtimes \mathcal{D}$ {is} likewise {a} modular tensor {category} over $\mathrm{sRep}(\mathrm{U}(1)^f)$. Suppose that the condensable algebra $\mathcal{A}$ contains a local boson $\mathbf{m}$. Because $\mathcal{A}$ must also include all objects generated by repeated fusion of $\mathbf{m}$, it necessarily contains 
\begin{align}
    \cdots, \bar{\mathbf{m}}, \mathbf{0}, \mathbf{m}, \mathbf{2m}, \mathbf{3m}, \cdots.  
\end{align} 
As a consequence, the M\"uger center \cite{lan2017classification,lan2017modular}, i.e., the subcategory formed by all the local objects, of the condensed theory is no longer $\mathrm{sRep}(\mathrm{U}(1)^f)$, but instead $\mathrm{sRep}\left(H^f\right)$ for some subgroup $H \subset \mathrm{U}(1)$. Since the M\"uger center characterizes the symmetry of the system \cite{lan2017classification,lan2017modular}, this can be physically understood as the symmetry breaking $\mathrm{U}(1) \to H$ \cite{lan2017classification,lan2017modular}. In particular, when $H=\mathbb{Z}_q$, i.e., when the smallest charged local boson $\mathbf{m}\neq \mathbf{0}$ in $\mathcal{A}$ is $\mathbf{q}$, the resulting theory describes a charge-$qe$ anyon superconductor. 

On the other hand, certain anyon condensations preserve $\mathrm{sRep}(\mathrm{U}(1)^f)$, so that the condensed theory remains a modular tensor category over $\mathrm{sRep}(\mathrm{U}(1)^f)$. In this case, the condensable algebra $\mathcal{A}$ contains no local bosons other than the vacuum $\mathbf{0}$. Consequently, the global $\mathrm{U}(1)$ symmetry is preserved, and the anyon condensation corresponds to a transition between two different $\mathrm{U}(1)$-symmetric topological orders. Since fusion preserves electric charge, the absence of charged local bosons in $\mathcal{A}$ implies that all anyons contained in $\mathcal{A}$ must be charge-neutral. This observation is consistent with Ref.~\cite{zhang2025hierarchy}, where the anyons in $\mathcal{A}$ are required to be charge-neutral in order to describe hierarchy constructions. Note that, in general, the anyon $a$ in the parent theory $\mathcal{C}$ has nonzero charge. Thus, for $a b$ to be charge-neutral to describe the hierarchy transition, $b$ should also have a nonzero charge. This implies that $\mathcal{D}$ is charged. As explained in Ref.~\cite{zhang2025hierarchy}, the Hall conductance of the hierarchical state is given by the sum of the Hall conductances of $\mathcal{C}$ and $\mathcal{D}$. Furthermore, the chiral central charge $c_{(\mathcal{C} \boxtimes \mathcal{D}) / \mathcal{A}}$ of the condensed theory is determined by the addition of those of $\mathcal{C}$ and $\mathcal{D}$~\cite{zhang2025hierarchy} 
\begin{align}
    c_{(\mathcal{C} \boxtimes \mathcal{D}) / \mathcal{A}} = c_{\mathcal{C}} + c_{\mathcal{D}}.  
\end{align}
The above relation for the chiral central charges also holds for anyon superconductors. 

We conclude with a practical remark. To carry out the procedure, one must identify a stacked topological order $\mathcal{D}$ containing an anyon $b$ such that, for a given anyon $a$ in the parent topological order $\mathcal{C}$, the composite $ab$ forms a condensable boson in the product theory $\mathcal{C}\boxtimes\mathcal{D}$. In general, however, there are infinitely many choices of $\mathcal{D}$ that satisfy this requirement, making additional guiding principles necessary. In this work, we adopt the following two criteria~\cite{zhang2025hierarchy}. First, we favor topological orders with smaller rank and smaller total quantum dimension. Second, we avoid topological orders with large chiral central charge, as stacking such theories would substantially modify the resulting chiral central charge without clear physical motivation.

\bgroup
\def\arraystretch{1.5}
\begin{table*}[t]
    \centering
    \caption{Examples of anyon superconductors and hierarchical quantum Hall states. Here $\mathcal{C}$ is the parent $\mathrm{U}(1)$-symmetric topological order, $a_q$ is the doped anyon, and $\mathcal{D}$ is the stacking theory. The anyon $a_q$ is specified by its quantum dimension $d$, topological spin $s$, and charge $q$. The final theory $(\mathcal{C} \boxtimes \mathcal{D}) / \mathcal{A}$ has the chiral central charge $c$. The unit charge of the system is denoted by $e$.}
    \begin{tabular*}{500pt}{@{\extracolsep{\fill}}cccccc}
        \hline \hline
        $\mathcal{C}$ & $a_0$ & $\mathcal{D}$ & $(\mathcal{C} \boxtimes \mathcal{D}) / \mathcal{A}$ & $c$ \\
        \hline
        Laughlin $\nu = \frac{1}{3}$ & $(d = 1, s = \frac{1}{3}, q = \frac{e}{3})$ & $\mathrm{U}(1)_{-2} \boxtimes \mathrm{U}(1)_6$ & Charge-$2e$ SC (Sec.~\ref{ex:fermionic_Laughlin}) & 1 \\
        Pfaffian & $(1, \frac{1}{4}, \frac{e}{2})$ & $\mathrm{U}(1)_{-2}$ & Charge-$2e$ SC (Sec.~\ref{ex:fermionic_Pfaffian}) & $\frac{1}{2}$ \\
        Pfaffian & $(1, -\frac{1}{4}, \frac{e}{2})$ & $\mathrm{U}(1)_2$ & Charge-$2e$ SC (Sec.~\ref{ex:fermionic_Pfaffian}) & $\frac{5}{2}$ \\
        $\mathbb{Z}_k$ Read-Rezayi & $(2 \cos \frac{\pi}{k + 2}, \frac{1}{2(k + 2)}, \frac{e}{k + 2})$ & $\left[\left(\mathrm{SU}(2)_k \boxtimes \mathrm{U}(1)_{-2(k + 2)}\right) \boxtimes \mathrm{SU}(2)_{-2}\right] / \mathbb{Z}_2$ & Charge-$2e$ SC (Sec.~\ref{ex:fermionic_Read-Rezayi}) & $-\frac{1}{2}$ \\
        $\mathbb{Z}_k$ Read-Rezayi (even $k$) & $(2 \cos \frac{\pi}{k + 2}, \frac{1}{2(k + 2)}, \frac{e}{k + 2})$ & $\mathrm{SU}(2)_{-k} \boxtimes \mathrm{U}(1)_{2 k} \boxtimes \mathrm{U}(1)_{-k (k + 2)}$ & Charge-$ke$ SC (Sec.~\ref{ex:fermionic_Read-Rezayi}) & $0$ \\ 
        \hline
        Laughlin $\nu = \frac{1}{2}$ & $(1, \frac{1}{4}, \frac{e}{2})$ & $\mathrm{U}(1)_{-2}$ & Charge-$e$ SC (Sec.~\ref{ex:bosonic_Laughlin}) & $0$ \\
        $\mathbb{Z}_k$ bosonic Read-Rezayi & $(2 \cos \frac{\pi}{k + 2}, \frac{3}{4(k + 2)}, \frac{e}{2})$ & $\mathrm{SU}(2)_{-k} \boxtimes D(\mathbb{Z}_{2k})$ & Charge-$ke$ SC (Sec.~\ref{ex:bosonic_Read-Rezayi_SC}) & $0$ \\
        $\mathbb{Z}_k$ bosonic Read-Rezayi & $(2 \cos \frac{\pi}{k + 2}, \frac{3}{4(k + 2)}, \frac{e}{2})$ & $\mathrm{SU}(2)_{-k} \boxtimes \mathrm{TC}$ & Toric code (Sec.~\ref{ex:bosonic_Read-Rezayi_TC}) & $0$ \\
        \hline \hline
    \end{tabular*}
    \label{tab:examples}
\end{table*}
\egroup

\subsection{Minimal stack-and-condense route to anyon superconductivity} \label{sec:minimal}

Interestingly, the above stack-and-condense operation immediately reveals an existence of the path to achieve charge-$2e$ anyon superconductivity in fermionic systems and charge-$e$ anyon superconductivity in bosonic systems through the doping of minimally charged anyons. For fermionic systems, this was also noted in Ref.~\cite{shi2025doping} within the conventional modular tensor category framework.

We first explain how to identify a stacking theory $\mathcal{D}$ and the associated condensable algebra $\mathcal{A}$ that give rise to a charge-$2e$ superconductor, starting from a given $\mathrm{U}(1)$-symmetric fermionic topological order $\mathcal{C}$. From $\mathcal{C}$, we first construct a supermodular tensor category $\mathcal{C}_0$ by condensing all local bosons in $\mathcal{C}$, namely $\mathcal{C}_0=\mathcal{C}/\mathcal{A}_\mathrm{local}$ with $\mathcal{A}_\mathrm{local} = \bigoplus_{\mathbf{n}\in 2\mathbb{Z}}\mathbf{n}$. We then take the time-reversal conjugate of $\mathcal{C}_0$, denoted by $\bar{\mathcal{C}}_0$. Finally, we compute a minimal modular extension of $\bar{\mathcal{C}}_0$, which we denote by $\bar{\mathcal{C}}_0^{(b)}$, and set $\mathcal{D}=\bar{\mathcal{C}}_0^{(b)}$. We note that the anyons of $\mathcal{D}$ are all charge-neutral, while their topological data are the time-reversed counterparts of those in $\mathcal{C}$, apart from the charge assignments. The above steps should be understood as a mathematical procedure for systematically constructing an appropriate stacking theory $\mathcal{D}$.

The stacked theory $\mathcal{C}\boxtimes\mathcal{D}$ then contains a canonical condensable algebra 
\begin{align}
\mathcal{A} = \bigoplus_{a \in \mathcal{C}} a_{q_a} \bar{a},
\end{align}
which is generated by the composite of a minimally charged anyon in $\mathcal{C}$ and its partner in $\mathcal{D}$. Here, we label anyons in $\mathcal{C}$ by a subscript $q_a$ to emphasize that they carry electric charge, whereas the anyons in $\mathcal{D}$ are charge-neutral. The anyons of $\mathcal{D}$ are denoted by $\bar a$, reflecting that they have topological data conjugate to those of $a_{q_a}\in\mathcal{C}$. For brevity, we refer to $\bar a$ as the time-reversal conjugate of $a_{q_a}$, with the understanding that $\bar a$ is charge-neutral. By construction, the condensable algebra $\mathcal{A}$ contains the local boson $\mathbf{2}$. Condensing $\mathcal{A}$ confines all nonlocal anyons, so that the resulting phase is a charge-$2e$ anyon superconductor with no residual topological order. Moreover, since the chiral central charge of the minimal modular extension $\bar{\mathcal{C}}_0^{(b)}$ takes the form $-c_{\mathcal{C}}+\tfrac{k}{2}$ with $k\in\mathbb{Z}$~\cite{lan2017classification}, the chiral central charge of the resulting superconducting phase is $\tfrac{k}{2}$.

When $\mathcal{C}$ is a bosonic theory, we similarly obtain a modular tensor category $\mathcal{C}_0$ by condensing all local bosons $\mathcal{A}_\mathrm{local} = \bigoplus_{\mathbf{n} \in \mathbb{Z}} \mathbf{n}$. Then, we simply take $\mathcal{D} = \bar{\mathcal{C}}_0$. Again, $\mathcal{C} \boxtimes \mathcal{D}$ has a canonical condenable algebra $\mathcal{A} = \bigoplus_{a \in \mathcal{C}} a_{q_a} \bar{a}$, where $\bar{a}$ are the charge-neutral time-reversal conjugates of $a_{q_a} \in \mathcal{C}$. The condensable algebra $\mathcal{A}$ confines all the anyons and now contains the local boson $\mathbf{1}$. Thus, the condensed theory is a charge-$e$ superconductor without residual topological order. Since the chiral central charge of $\bar{\mathcal{C}}_0$ takes the form of $-c_\mathcal{C} + 8k$ with $k \in \mathbb{Z}$ \cite{wen2016theory}, the resulting superconductor has $c = 8k$.

\section{Fermionic Examples} \label{sec:fermionic}

We now present several examples illustrating our categorical descriptions of anyon superconductors in fermionic systems. The results are summarized in Tab.~\ref{tab:examples}. 

\subsection{Charge-\texorpdfstring{$2e$}{} superconductivity from Laughlin state} \label{ex:fermionic_Laughlin}

We revisit the superconductor derived from the Laughlin state in Sec.~\ref{sec:Laughlin} from a categorical perspective. The anyons of the Laughlin state are generated by an order-$3$ Abelian anyon $a_{\frac{1}{3}}$. As we discussed in Sec.~\ref{sec:Laughlin} within the field theoretic analysis, the anyon superconducor can emerge by stacking charge-neutral $\mathcal{D} = \mathrm{U}(1)_{-2} \times \mathrm{U}(1)_6$. Let $b$ and $c$ denote the generating anyons of $\mathrm{U}(1)_{-2}$ and $\mathrm{U}(1)_6$, respectively. Then, the condensable algebra is 
\begin{align}
    \mathcal{A} &= \cdots \oplus \mathbf{0} \oplus a_{\frac{1}{3}} b c \oplus a_{\frac{2}{3}} c^2 \oplus \mathbf{1} c^3 \nonumber \\
    &\quad \oplus \mathbf{1} a_{\frac{1}{3}} b c^4 \oplus \mathbf{1} a_{\frac{2}{3}} c^5 \oplus \mathbf{2} \oplus \cdots.
\end{align}
The resulting state is a charge-$2e$ anyon superconductor with $c = 1$. This aligns with the field-theoretic description in Sec.~\ref{sec:Laughlin}.
 
\subsection{Charge-\texorpdfstring{$2e$}{} superconductivity from Pfaffian state with nonminimally charged anyons} \label{ex:fermionic_Pfaffian}

Since charge-$2e$ superconductors obtained by doping minimally charged anyons admit a relatively straightforward categorical description (Sec.~\ref{sec:minimal}), here we consider an example in which superconductivity arises from doping a nonminimally charged Abelian anyon in the Pfaffian state, as analyzed in Ref.~\cite{shi2025doping}.

The Pfaffian state contains five nontrivial anyons: $\sigma_{\frac{1}{4}}$, $\psi_0$, $\alpha_{\frac{1}{2}}$, $\bar{\sigma}_{\frac{3}{4}}$, and $\bar{\alpha}_{\frac{1}{2}}$. We consider doping the Abelian anyon $\alpha_{\frac{1}{2}}$ with the semionic self-statistics. For this, we stack the theory with charge-neutral $\mathcal{D} = \mathrm{U}(1)_{-2}$, which contains the anti-semion $\bar{s}$. We then condense the algebra
\begin{align}
    \mathcal{A} &= \bigoplus_{\mathbf{m} \in 2 \mathbb{Z}} \mathbf{m} \left(\mathbf{0} \oplus \alpha_{\frac{1}{2}} \bar{s} \oplus \mathbf{1} \psi_0 \oplus \mathbf{1} \bar{\alpha}_{\frac{1}{2}} \bar{s}\right) \nonumber \\
    &= \cdots \oplus \mathbf{0} \oplus \alpha_{\frac{1}{2}} \bar{s} \oplus \mathbf{1} \psi_0 \oplus \mathbf{1} \bar{\alpha}_{\frac{1}{2}} \bar{s} \oplus \mathbf{2} \oplus \cdots,
\end{align}
which is generated by the composite anyon $\alpha_{\frac{1}{2}} \bar{s}$. Since $\mathcal{A}$ contains the local boson $\mathbf{2}$ and completely confines all the nonlocal anyons, the resulting phase is a charge-$2e$ anyon superconductor with no residual topological order. The chiral central charge of this superconductor is given by the sum of those of the Pfaffian state and $\mathcal{D}$, yielding $c = \frac{3}{2} - 1 = \frac{1}{2}$. This result is in agreement with Ref.~\cite{shi2025doping}, whose field-theoretic description of this superconductor is reviewed in Appendix~\ref{app:review}.

Next, we consider doping the anti-semionic Abelian anyon $\bar{\alpha}_{\frac{1}{2}}$. In this case, we stack the theory with $\mathcal{D} = \mathrm{U}(1)_2$, which contains the semion $s$, and proceed to condense
\begin{align}
    \mathcal{A} &= \bigoplus_{\mathbf{m} \in 2 \mathbb{Z}} \mathbf{m} \left(\mathbf{0} \oplus \Bar{\alpha}_{\frac{1}{2}} s \oplus \mathbf{1} \psi_0 \oplus \mathbf{1} \Bar{\alpha}_{\frac{1}{2}}\right) \nonumber \\
    &= \cdots \oplus \mathbf{0} \oplus \Bar{\alpha}_{\frac{1}{2}} s \oplus \mathbf{1} \psi_0 \oplus \mathbf{1} \Bar{\alpha}_{\frac{1}{2}} s \oplus \mathbf{2} \oplus \cdots,
\end{align}
which is generated by $\Bar{\alpha}_{\frac{1}{2}} s$. The resulting phase is a charge-$2e$ superconductor with chiral central charge $c=\tfrac{5}{2}$. This result is consistent with the corresponding field-theoretic analysis {of Ref.}~\cite{shi2025doping}{, which is reviewed in Appendix~\ref{app:review}}.

\subsection{Charge-\texorpdfstring{$ke$}{} superconductivity from \texorpdfstring{$\mathbb{Z}_k$}{} Read-Rezayi state} \label{ex:fermionic_Read-Rezayi}

We consider doping the non-Abelian anyon $a_{\frac{1}{k + 2}}$ of the $\mathbb{Z}_k$ Read-Rezayi state, which has quantum dimension $d = 2 \cos(\tfrac{\pi}{k + 2})$, topological spin $s = \frac{1}{2 (k + 2)}$, and electric charge $q = \frac{e}{k + 2}$. For even $k$, {if} $a_{\frac{1}{k + 2}}$ energetically favors the ferromagnetic fusion channel, the field-theoretic analysis of Ref.~\cite{shi2025doping} predicts that the doped system realizes charge-$ke$ superconductivity. This field theory is reviewed in Appendix~\ref{app:review} for completeness. 

We now construct the stack-and-condense formulation of the charge-$ke$ superconductor. 
To this end, we stack the following charge-neutral theory 
\begin{align}
\mathcal{D} = \mathrm{SU}(2)_{-k} \times \mathrm{U}(1)_{2 k} \times \mathrm{U}(1)_{-k (k + 2)}  
\end{align}
onto the $\mathbb{Z}_k$ Read-Rezayi state. We denote the $j = \tfrac{1}{2}$ anyon of $\mathrm{SU}(2)_{-k}$ by $\bar{\sigma}$, and the generators of $\mathrm{U}(1)_{2k}$ and $\mathrm{U}(1)_{-k(k+2)}$ by $b$ and $\bar{c}$, respectively. The condensable algebra $\mathcal{A}$ is then generated by the composite anyon $a_{\frac{1}{k + 2}} \bar{\sigma} b \bar{c}$. Since $\bar{\sigma} b \bar{c}$ has order $k(k+2)$ and $a_{\frac{1}{k + 2}}$ carries charge $e/(k+2)$, the local bosons appearing in $\mathcal{A}$ are generated by charge $ke$. Consequently, condensing $\mathcal{A}$ yields a charge-$ke$ superconductor. Furthermore, this condensation leaves no residual topological order. The chiral central charge of the resulting phase is $c=0$, as the contributions from $\mathcal{C}$ and $\mathcal{D}$ exactly cancel. All these features are consistent with the field-theoretic analysis~\cite{shi2025doping}. 

As a concrete example, let us present the $\mathbb{Z}_4$ Read-Rezayi state where $a_{\frac{1}{6}}$ is doped and favors the ferromagnetic fusion channel. The anyons of the $\mathbb{Z}_4$ Read-Rezayi state are labeled by pairs $(j,n)$, where $j=\tfrac{j'}{2}$ with $0\le j'\le 4$ and $0\le n\le 12$, subject to the constraint $j'+n\in 2\mathbb{Z}$. In this notation, the non-Abelian anyon $a_{\frac{1}{6}}$ corresponds to $(\tfrac{1}{2},1)$, while the electron $\mathbf{1}$ is given by $(2,6)$. Following the prescription above, we stack the charge-neutral
\begin{align}
\mathcal{D} = \mathrm{SU}(2)_{-4} \times \mathrm{U}(1)_8 \times \mathrm{U}(1)_{-24}.  
\end{align}
We label the anyons of $\mathcal{D}$ by $(\bar{j}, b^l, c^l)$, where $\bar{j}=\tfrac{\bar{j}'}{2}$ with $0\le \bar{j}'\le 4$, and $b$ and $c$ generate the $\mathrm{U}(1)_8$ and $\mathrm{U}(1)_{-24}$ sectors, respectively. We then condense 
\begin{align}
    \mathcal{A} &= \cdots \oplus \mathbf{0} \oplus (1, 0) (\bar{1}, b^0, c^0) \oplus (2, 0) (\bar{2}, b^0, c^0) \nonumber \\
    &\quad \oplus \left(\frac{1}{2}, 1\right) \left(\bar{\frac{1}{2}}, b, c\right) \oplus \left(\frac{3}{2}, 1\right) \left(\bar{\frac{3}{2}}, b, c\right) \oplus \cdots \nonumber \\
    &\quad \cdots \oplus \mathbf{3} \left(\frac{3}{2}, 5\right) \left(\bar{\frac{3}{2}}, b^7, c^{23}\right) \oplus \mathbf{4} \oplus \cdots.
\end{align}
Note that $\mathcal{A}$ is generated by the composite anyon $\left(\tfrac{1}{2},1\right)\left(\bar{\tfrac{1}{2}},b,c\right)$, in the sense that all other anyons in $\mathcal{A}$ can be obtained by repeated fusion of this generator and its anti-anyon. Since all local bosons contained in $\mathcal{A}$ are generated by $\mathbf{4}$, and condensing $\mathcal{A}$ completely trivializes the topological order, the resulting phase is a charge-$4e$ superconductor with no residual topological order. This result is consistent with the field-theoretic analysis~\cite{shi2025doping}.

This finding is also consistent with Ref.~\cite{lan2017classification}, which showed that all eight minimal modular extensions of $\mathrm{sRep}(\mathbb{Z}_4^f)$ are Abelian and carry integer-valued chiral central charge. Physically, this implies that charge-$4e$ superconductors arising from fermionic systems cannot support non-Abelian quasiparticles and must have integer-valued chiral central charge. Indeed, the charge-$4e$ anyon superconductor discussed above hosts no non-Abelian excitations and has an integer chiral central charge, $c = 0$.

\section{Bosonic Examples} \label{sec:bosonic}

We now turn to bosonic examples, demonstrating the versatility of our approach in both fermionic and bosonic systems.

\subsection{Charge-\texorpdfstring{$e$}{} superconductivity from bosonic Laughlin state} \label{ex:bosonic_Laughlin}

Our first bosonic example is the Laughlin state at filling fraction $\nu=\tfrac{1}{2}$, which realizes the semion topological order. The semion carries electric charge $e/2$. Field-theoretic analysis~\cite{shi2025doping} shows that doping the semion yields a charge-$e$ superconductor with vanishing chiral central charge. For completeness, this analysis is reviewed in Appendix~\ref{app:review}.

In our categorical formalism, the Laughlin state contains an order-2 semion $s_{\frac{1}{2}}$ satisfying $s_{\frac{1}{2}} \otimes s_{\frac{1}{2}} = \mathbf{1}$, which is a local boson. To obtain the superconductor, we stack the theory with charge-neutral $\mathcal{D} = \mathrm{U}(1)_{-2}$ and condense the algebra
\begin{align}
    \mathcal{A} = \bigoplus_{\mathbf{n} \in \mathbb{Z}} \bigoplus_{l=0}^1 \mathbf{n} s_{\frac{1}{2}}^l \bar{s}^l,  
\end{align}
whose condensation completely trivializes the topological order. Moreover, $\mathcal{A}$ contains local bosons carrying arbitrary integer charge, thereby fully breaking the global $\mathrm{U}(1)$ charge conservation. Consequently, the resulting phase is a charge-$e$ superconductor without residual topological order, consistent with the field-theoretic analysis~\cite{shi2025doping}.

\subsection{Charge-\texorpdfstring{$ke$}{} superconductivity from bosonic \texorpdfstring{$\mathbb{Z}_k$}{} Read-Rezayi state} \label{ex:bosonic_Read-Rezayi_SC}

We now turn to the bosonic analogue of charge-$ke$ anyon superconductivity in electronic systems (Sec.~\ref{sec:fermionic}), in which distinct phases emerge upon doping a minimally charged non-Abelian anyon. Here we focus on the case where the ferromagnetic fusion channel of the minimally charged anyon is energetically favored, and we analyze this problem from both field-theoretic and categorical perspectives.

Technically speaking, the bosonic $\mathbb{Z}_k$ Read-Rezayi state is effectively described by \cite{seiberg2016gapped}
\begin{align} \label{eq:bRR_Lagrangian}
    \mathcal{L}_{\mathrm{bRR}_k} &= - \frac{k}{4 \pi} \Tr\left[c d c + \frac{2}{3} c^3\right] + \frac{k}{4 \pi} (\Tr c) d (\Tr c) \nonumber \\
    &\quad + \frac{1}{2 \pi} (\Tr c) d b + \frac{1}{2 \pi} A d b + \mathcal{L}[\Phi, c]
\end{align}
where $b$ and $c$ are dynamical $\mathrm{U}(1)$ and $\mathrm{U}(2)$ gauge fields, respectively, and $\Phi$ transforms in the fundamental representation of $\mathrm{U}(2)$. We consider doping the minimal non-Abelian anyon $a_{\frac{1}{2}} = (\frac{1}{2}, 1)$ to the system. For the notation $(j,n)$ and anyon contents of the theory, we refer to App.~\ref{app:data}.

We first consider the ferromagnetic case to obtain charge-$ke$ superconductor, in which the fusion of the minimally charged non-Abelian anyons $a_{\frac{1}{2}}=\left(\tfrac{1}{2},1\right)$ energetically favors the triplet channel, leading to spontaneous symmetry breaking of $\mathrm{U}(2)$ down to $\mathrm{U}(1)\times\mathrm{U}(1)$~\cite{shi2025doping}. This can be effectively encoded by setting $c \to \operatorname{diag}(c_\uparrow, c_\downarrow)$ 
\begin{align}
    \mathcal{L}_{\mathrm{bRR}_k} &\to \frac{k}{2 \pi} c_\uparrow d c_\downarrow + \frac{1}{2 \pi} b d (c_\uparrow + c_\downarrow) \nonumber \\
    &\quad + \frac{1}{2 \pi} A d b + \mathcal{L}[\Phi, c_\uparrow]. 
\end{align}
The equation of motion for $c_\uparrow$ implies that $\Phi$ is at the effective filling $\nu = 2 k$. We assume that $\Phi$ forms $k$ copies of the bosonic integer quantum Hall states, yielding 
\begin{align}
    \mathcal{L}[\Phi, c_\uparrow] \to \frac{2 k}{4 \pi} c_\uparrow d c_\uparrow.
\end{align}
In the $K$-matrix formulation (App.~\ref{app:K-matrix}), the resulting theory is represented as 
\begin{align}
    K = 
    \begin{pmatrix}
        -2k & -k & -1 \\
        -k & 0 & -1 \\
        -1 & -1 & 0
    \end{pmatrix}, \quad 
    \mathbf{t} = 
    \begin{pmatrix}
        0 \\ 0 \\ 1
    \end{pmatrix}.
\end{align}
We then perform the following basis transformation: 
\begin{align}
    W^\mathsf{T} &= 
    \begin{pmatrix}
        1 & 0 & 0 \\
        0 & 1 & k-1 \\
        -1 & 1 & k
    \end{pmatrix},  \\
    W^\mathsf{T} K W &= 
    \begin{pmatrix}
        -2k & 1 - 2k & 0 \\
        1 - 2k & 2 - 2k & 0 \\ 
        0 & 0 & 0
    \end{pmatrix}, \quad 
    W^\mathsf{T} \mathbf{t} = 
    \begin{pmatrix}
        0 \\ k - 1 \\ k
    \end{pmatrix}.
\end{align}
Integrating out the upper $2 \times 2$ block, we finally get 
\begin{align}
    \mathcal{L}_\mathrm{SC} = \frac{k}{2 \pi} A d a + \frac{2k (k - 1)^2}{4 \pi} A d A.
\end{align}
This describes a charge-$ke$ superconductor with vanishing chiral central charge.

This anyon superconductivity can also be derived within our categorical formulation, providing a nontrivial consistency check against the field-theoretic analysis. Since we are dealing with bosonic systems, we work with modular tensor categories over $\mathrm{Rep}\left(\mathrm{U}(1)\right)$. The anyons of the bosonic $\mathbb{Z}_k$ Read-Rezayi state are described by the $\mathrm{SU}(2)_k$ modular tensor category, whose data is in App.~\ref{app:data}. The doped anyon $a_{\frac{1}{2}}$ corresponds to the $j = \frac{1}{2}$ representation of $\mathrm{SU}(2)_k$. Let $\mathcal{D}=\mathrm{SU}(2)_{-k}\boxtimes D(\mathbb{Z}_{2k})$, where $D(\mathbb{Z}_{2k})$ denotes the quantum double of $\mathbb{Z}_{2k}$, i.e., the $\mathbb{Z}_{2k}$ gauge theory (its anyon content is summarized in App.~\ref{app:data}). Furthermore, we set $\mathcal{D}$ charge-neutral and thus its anyons do not carry electric charges. We will denote the $\mathrm{SU}(2)_{-k}$ anyons in $\mathcal{D}$ as $\bar{a}$. The condensable algebra is then 
\begin{align} \label{eq:condensable_bosonic_RR_SC}
    \mathcal{A} = \bigoplus_{\mathbf{n} \in k\mathbb{Z}} \bigoplus_{l = 0}^{2k - 1} \bigoplus_{a \in \mathrm{SU}(2)_k} \mathbf{n} a_{q_a} \bar{a} e^l
\end{align}
where $e$ is the charge-neutral, minimal electric excitation of the $\mathbb{Z}_{2k}$ gauge theory $D(\mathbb{Z}_{2k})$, which has spin $0$ and order $2k$. Note that the condensable algebra $\mathcal{A}$ is generated by repeated fusion of the composite anyon $a_{\frac{1}{2}}...$ with the doped minimal anyon $a_{\frac{1}{2}}$ and contains local bosons generated by $\mathbf{k}$. Moreover, $\mathcal{D}$ carries chiral central charge opposite to that of the bosonic $\mathbb{Z}_k$ Read-Rezayi state. Consequently, the condensed phase is a charge-$ke$ superconductor with vanishing chiral central charge, $c=0$. This is consistent with the field-theoretic analysis presented above.

\subsection{Toric code from bosonic \texorpdfstring{$\mathbb{Z}_k$}{} Read-Rezayi state} \label{ex:bosonic_Read-Rezayi_TC}

We next consider the paramagnetic channel, in which the $\mathrm{U}(2)$ gauge symmetry remains fully preserved. In this case, the equation of motion for the gauge field $c$ constrains the scalar field $\Phi$ to an effective filling $\nu=-2k$. Thus, we assume that $\Phi$ realizes $k$ copies of the bosonic integer quantum Hall state at filling $\nu=-2$ 
\begin{align} \label{eq:bosonic_RR_k_matter_field}
    \mathcal{L}[\Phi, c] \to \frac{k}{4 \pi} \Tr\left[c d c + \frac{2}{3} c^3\right] - \frac{k}{4 \pi} (\Tr c) d (\Tr c).
\end{align}
As a result, we obtain
\begin{align} \label{eq:Pfaffian_toric_code}
    \mathcal{L}_\mathrm{bRR_k} \rightarrow \mathcal{L}_\mathrm{TC} = \frac{1}{2 \pi} (\Tr c) db + \frac{1}{2 \pi} A d b.
\end{align}
Expressing $\Tr c = 2 c_0$ where $c_0$ is a $\mathrm{U}(1)$ gauge field, we then see that Eq.~\eqref{eq:Pfaffian_toric_code} is the toric code topological order. Let the anyons in this toric code state be denoted by $\{e, m, f = e \otimes m\}$. Then, the coupling term with $A$ in Eq.~\eqref{eq:Pfaffian_toric_code} implies that $e$ and $f$ carries charge $-\frac{1}{2}$ while $m$ is charge-neutral. Notably, doping minimally charged anyons that favor the paramagnetic fusion channel does not lead to superconductivity; instead, it preserves the global charge conservation and drives a quantum Hall hierarchy transition to another topological order.

Within our categorical approach, we stack \textit{charged} $\mathcal{D} = \mathrm{SU}(2)_{-k} \boxtimes \mathrm{TC}$, in which $\mathrm{TC}$ represents the toric code topological order. We also set the anyons of $\mathrm{SU}(2)_{-k}$ carry exactly the opposite charges to the bosonic $\mathbb{Z}_k$ Read-Rezayi state. The anyons in $\mathrm{TC}$ is assumed to have charges $q_e = q_f = -\frac{1}{2}$ and $q_m = 0$. We then condense 
\begin{align}
    \mathcal{A} = \bigoplus_{a_q \in \mathrm{SU}(2)_k} a_q \bar{a}_{-q}.
\end{align}
Since we consider charged $\mathcal{D}$ where the charges of its $\mathrm{SU}(2)_{-k}$ part exactly cancel the charges of the bosonic $\mathbb{Z}_k$ Read-Rezayi state, all the condensing anyons are charge-neutral. Thus, the condensation of $\mathcal{A}$ preserves the global $\mathrm{U}(1)$ symmetry and the system. After the condensation, the $\mathrm{SU}(2)_k \boxtimes \mathrm{SU}(2)_{-k}$ part is trivialized and the remaining topological order is $\mathrm{TC}$, in agreement with our field-theoretic analysis. 

\section{Conclusion} \label{sec:conclusion}

In this work, we developed a categorical framework that unifies quantum Hall hierarchy constructions and anyon superconductivity. Conserved $\mathrm{U}(1)$ charge is tracked using modular tensor categories over $\mathrm{Rep}(\mathrm{U}(1))$ or $\mathrm{sRep}(\mathrm{U}(1)^f)$, where local integer-charged excitations form the M\"uger center. Doping a quantum Hall state $\mathcal{C}$ is captured by stacking an auxiliary topological order $\mathcal{D}$ and condensing a condensable algebra $\mathcal{A}$ generated by a composite anyon $ab$, with $a\in\mathcal{C}$ and $b\in\mathcal{D}$. When $\mathcal{A}$ contains charged local bosons, the M\"uger center is broken to its discrete subgroup and the system enters an anyon superconducting phase. On the other hand, when all anyons in $\mathcal{A}$ are charge-neutral, the M\"uger center is preserved and the transition yields a hierarchy quantum Hall state. This framework reproduces all known examples of anyon superconductivity~\cite{shi2024doping,shi2025doping} and predicts new fermionic and bosonic anyon superconductors, with the condensate charge fixed unambiguously by the charged local bosons in $\mathcal{A}$. Looking ahead, our results suggest a systematic route to classifying anyon superconductivity emerging from fractional Chern insulators, an avenue that merits further investigation. 

\begin{acknowledgments}
This work is financially supported by Samsung Science and Technology Foundation under Project Number SSTF-BA2401-03, the NRF of Korea (Grants No. RS-2024-00410027, RS-2023-NR119931, RS-2024-00444725, RS-2023-00256050, IRS-2025-25453111, RS-2025-08542968) funded by the Korean Government (MSIT), the Air Force Office of Scientific Research under Award No. FA23862514026, and Institute of Basic Science under project code IBS-R014-D1. T.~L. is partially supported by KAIST Undergraduate Research Program (URP)
\end{acknowledgments}

\appendix

\section{\texorpdfstring{$K$}{K}-matrix formulation} \label{app:K-matrix}

Every $(2 + 1)$-dimensional Abelian topological order with $\mathrm{U}(1)$ symmetry is characterized by the $K$-matrix and the charge vector $\mathbf{t}$ \cite{wen1992classification}. The effective Lagrangian of an arbitrary Abelian topological order can be written as 
\begin{align}
    \mathcal{L} = -\frac{K_{IJ}}{4 \pi} a^I d a^J + \frac{t_I}{2\pi} a^I d A
\end{align}
where $a^I$ with $I = 1, \cdots, N$ are $\mathrm{U}(1)$ gauge fields and $A$ is the background electromagnetic gauge potential. Here, $K$ is an $N \times N$ nondegenerate, symmetric integer matrix and $\mathbf{t}$ is an $N$-dimensional integer vector. The Hall conductance is then given by $\sigma_H = \mathbf{t}^\mathsf{T} K^{-1} \mathbf{t}$. When the given system obeys the spin-charge relation, i.e., the microscopic constituents of the system are fermions with unit charge, then the constraints $K_{II} = t_I$ mod $2$ are imposed \cite{seiberg2016gapped}.

\section{Details of the Laughlin state example} \label{app:details}

The fact that Eq.~\eqref{eq:Laughlin_anyon_superconductor_K} indeed describes a charge-$2e$ anyon superconductor can be shown by performing the following basis transformation:
\begin{align}
   W^\mathsf{T} &= 
   \begin{pmatrix}
       1 & 0 & 0 & 0 \\
       0 & 1 & 0 & 0 \\
       0 & 0 & 1 & 0 \\
       2 & 6 & -3 & 1
   \end{pmatrix},  \\
   W^\mathsf{T} K W &= 
   \begin{pmatrix}
       3 & -1 & 0 & 0 \\
       -1 & 0 & -1 & 0 \\
       0 & -1 & -2 & 0 \\
       0 & 0 & 0 & 0
   \end{pmatrix}, \quad 
   W^\mathsf{T} \mathbf{t} = 
   \begin{pmatrix}
       1 \\ 0 \\ 0 \\ 2
   \end{pmatrix}.
\end{align}
One can easily see that the upper $3 \times 3$ block of $W^\mathsf{T} K W$ describes the $\nu = 1$ integer quantum Hall state. By integrating out the associated dynamical gauge fields, the upper block gives rise to $\frac{1}{4 \pi} A d A + 2 \mathrm{CS}_g$ which contributes the chiral central charge $c = 1$. After the transformation, we are left with a $\mathrm{U}(1)$ gauge field $\tilde{a}$ without self Chern-Simons term, which couples to $A$ with the level-$2$ mutual Chern-Simons term. Then, the monopoles of $\tilde{a}$ condense, giving rise to the charge-$2e$ superconductivity \cite{seiberg2016gapped,ma2020emergent}.

\section{Brief introduction to algebraic theory of anyons} \label{app:mtc}

In this section, we provides an introduction to modular tensor category which serves as an algebraic theory of anyons \cite{kitaev2006anyons}. We will focus on physical intuition and encourage readers to refer to Ref.~\cite{etingof2015tensor} for further mathematical details.

In $2 + 1$ dimensions, topological orders are classified by their anyonic structures and chircal central charges \cite{kitaev2006anyons,wen2016theory,lan2016theory,lan2017classification,bruillard2017fermionic,ng2023reconstruction,cho2023classification,seo2024modular,ng2025classification}. Such an understanding not only allows for a systematic classification \cite{wen2016theory,lan2016theory,lan2017classification,ng2023reconstruction,cho2023classification,seo2024modular,ng2025classification} of topological orders but also provides a gateway to explore various physical properties of topological orders, including the gappability of edge \cite{levin2013protected,you2024gapped}, time-reversal symmetry \cite{geiko2024when}, and classical simulability \cite{ringel2017quantized,smith2020intrinsic,golan2020intrinsic,seo2025most}. 

For bosonic topological orders, the anyons form a mathematical structure known as a (unitary) modular tensor category \cite{kitaev2006anyons,wen2016theory,ng2023reconstruction,ng2025classification}. Roughly speaking, a modular tensor catogory $\mathcal{C}$ is a set of anyon types, called objects, endowed with the fusion rules and statistics of the anyons. Let $a, b, c \in \mathcal{C}$ then the fusion rules are given by 
\begin{align}
    a \otimes b = \bigoplus_{c \in \mathcal{C}} N^{ab}_c c,
\end{align}
where the nonnegative integers $N^{ab}_c$ are called fusion coefficients. There always exists the vacuum or the trivial anyon $\mathbf{0} \in \mathcal{C}$, which serves as the identity element under fusion. The statistics of the anyons are encoded in so-called modular data, a pair of matrices $(S, T)$. More precisely, $S$ is a symmetric unitary matrix encoding the mutual statistics of the anyons. The quantum dimension $d_a$ of an anyon $a$, which specifies the growth of the Hilbert space dimension per a creation of $a$, is given by $d_a = \frac{S_{\mathbf{0} a}}{S_{\mathbf{0} \mathbf{0}}}$. Meanwhile, $T$ is a diagonal unitary matrix encoding the self-statistics of anyons, where its diagonal elements are given by $T_{aa} = e^{2 \pi i s_a}$. Here, $s_a$ is called the topological spin of anyon $a$.

For fermionic topological order, there exists a fermionic local excitation which braids trivially with all other excitations. To encode this, we add a local object $\psi$ called the fundamental fermion, which has quantum dimension $d_\psi = 1$ and topological spin $s_\psi = \frac{1}{2}$. Such a mathematical structure is known as a supermodular tensor category \cite{lan2016theory,lan2017classification,bruillard2017fermionic,cho2023classification,seo2024modular}. Since $\psi$ braids trivially with other objects, $S$ is now nonunitary and the modular data admits the decomposition 
\begin{align} \label{eq:decomposition}
    S = \hat{S} \otimes \frac{1}{\sqrt{2}} 
    \begin{pmatrix}
        1 & 1 \\
        1 & 1
    \end{pmatrix}, \quad 
    T = \hat{T} \otimes 
    \begin{pmatrix}
        1 & 0 \\
        0 & -1
    \end{pmatrix}.
\end{align}
A supermodular tensor category $\mathcal{C}$ always admits a faithful embedding to a modular tensor category $\mathcal{C} \hookrightarrow \mathcal{C}^{(b)}$, known as the minimal modular extension \cite{lan2016theory,lan2017classification,lan2017modular,bruillard2017fermionic,johnson-freyd2024minimal,seo2024modular}. Moreover, $\mathcal{C}$ always has sixteen different minimal modular extensions which have chiral central charges differing by $\frac{1}{2}$. Physically, the minimal modular extension amounts to gauging the $\mathbb{Z}_2$ fermion-parity symmetry \cite{lan2016theory,lan2017classification,lan2017modular,bruillard2017fermionic,johnson-freyd2024minimal,seo2024modular}.

\section{Categorical description of topological orders with symmetry} \label{app:categorical}

In this section, we will briefly explain how the formalism of modular tensor category is generalized for topological orders with an onsite finite symmetry. Further details can be found in Refs.~\cite{lan2017classification,lan2017modular}.

When a given topological order has an onsite finite symmetry $G$, then the system hosts local quasiparticles carrying representations of $G$ on top of the nonlocal anyons. These local quasiparticles form a symmetric fusion category $\mathrm{Rep}(G)$ or $\mathrm{sRep}(G^f)$, depending on whether the system is bosonic or fermionic \cite{lan2017classification}. Here, $G^f$ is defined as the pair $(G, z)$, where $z \in G$ is an order-$2$ group element, i.e., $z \cdot z = E$ where $E$ is the identity element. In $\mathrm{sRep}(G^f)$, the representations which represents $z$ nontrivially has the fermionic self-statistics \cite{lan2017classification}. For example, consider $\mathrm{sRep}(\mathrm{U}(1)^f)$ which is relevant to our discussion. For $G = \mathrm{U}(1)$, the order-$2$ group element is unique $z = e^{i \pi}$. Since the irreducible representations of $\mathrm{U}(1)$ are labeled by an integer $n$ such that $\rho_n: e^{i \theta} \mapsto e^{n i \theta}$, one can see that odd-integer irreducible representations corresponds to local fermions.

\section{Anyon condensation} \label{app:anyon}

In this section, we briefly review the formalism of bosonic anyon condensation, which was first introduced for modular tensor categories \cite{kong2014anyon,eliens2014diagrammatics,neupert2016boson,burnell2018anyon} and recently generalized to supermodular tensor categories \cite{zhang2025hierarchy} via minimal modular extension \cite{bruillard2017fermionic,lan2017modular,seo2024modular}.

For a given bosonic topological order $\mathcal{C}$, each potential anyon condensation is characterized by a condensable algebra, or a connected \'etale algebra $\mathcal{A} \in \mathcal{C}$ \cite{kong2014anyon}. The condensable algebra is given by $\mathcal{A} = \bigoplus_{a \in \mathcal{C}} m_a a$ where $m_a$ are nonnegative integer coefficients such that $m_\mathbf{0} = 1$. When the condensed theory is the trivial gapped phase, $\mathcal{A}$ is called Lagrangian algebra. Though the condensable algebras are mathematically well-defined \cite{kong2014anyon}, it is often difficult to prove that a given candidate $\mathcal{A}$ is indeed a valid condensable algebra. Thus, the following necessary but not sufficient conditions for condensable algebras are used in most practical applications \cite{eliens2014diagrammatics,neupert2016boson,zhang2025hierarchy}:
\begin{align} \label{eq:necessary_bosonic}
    S_\mathcal{C} \mathbf{A} = \mathbf{A} S_{\mathcal{C} / \mathcal{A}}, \quad T_\mathcal{C} \mathbf{A} = \mathbf{A} T_{\mathcal{C} / \mathcal{A}}.
\end{align}
Here, $(S_\mathcal{C}, T_\mathcal{C})$ and $(S_{\mathcal{C} / \mathcal{A}}, T_{\mathcal{C} / \mathcal{A}})$ are the modular data of the topological orders before and after the condensation. The integer matrix $\mathbf{A}$ is defined via the Lagrangian algebra of $\mathcal{C} \boxtimes \overline{\mathcal{C} / \mathcal{A}}$ as
\begin{align}
    \mathcal{A}_{\mathcal{C} \boxtimes \overline{\mathcal{C} / \mathcal{A}}} = \bigoplus_{a \in \mathcal{C}, b \in \overline{\mathcal{C} / \mathcal{A}}} A_{a b} a b.
\end{align}

For fermionic topological orders, the theory of condensable algebra has not been fully established. However, the above necessary conditions for bosonic topological orders can be exploited via minimal modular extension \cite{zhang2025hierarchy}. Recall that the modular data of a supermodular tensor category admits the decomposition Eq.~\eqref{eq:decomposition}. Then, the necessary but not sufficient conditions are written as \cite{zhang2025hierarchy}
\begin{align} \label{eq:necessary_fermionic}
    \hat{S}_\mathcal{C} \mathbf{A} = \mathbf{A} \hat{S}_{\mathcal{C} / \mathcal{A}}, \quad \hat{T}^2_\mathcal{C} \mathbf{A} = \mathbf{A} \hat{T}^2_{\mathcal{C} / \mathcal{A}}.
\end{align}

In the main text, we presented a generalization of the conventional condensable algebra to modular tensor categories over $\mathrm{Rep}(\mathrm{U}(1))$ or $\mathrm{sRep}(\mathrm{U}(1)^f)$ with infinitely many objects, for which we cannot directly use the above necessary conditions of condensable algebra. Let $\mathcal{C}$ be a given modular tensor category over $\mathrm{Rep}(\mathrm{U}(1))$ or $\mathrm{sRep}(\mathrm{U}(1)^f)$. Then, we first obtain the ``base'' category $\mathcal{C}_0$ by condensing all the local bosons of $\mathcal{C}$. Then, $\mathcal{C}_0$ would be either a modular tensor category or a supermodular tensor category with well-defined modular data, allowing a condensable algebra $\mathcal{A}_0$ to be found via the above necessary conditions Eqs.~\eqref{eq:necessary_bosonic} or \eqref{eq:necessary_fermionic}. Then, the condensable algebra of $\mathcal{C}$ takes the form of $\mathcal{A} = \bigoplus_{\mathbf{n} \in k \mathbb{Z}} \mathbf{n} \mathcal{A}_0$ for an integer $k$, if anyons in $\mathcal{A}_0$ are charged. When all the anyons in $\mathcal{A}_0$ are charge-neutral, the condensable algebra is simply given by $\mathcal{A} = \mathcal{A}_0$.

\section{Review on field-theoretic derivation of anyon superconductors} \label{app:review}

For completeness of our manunscript, we review the field-theoretic derivation of anyon superconductors presented in literature \cite{shi2024doping,shi2025doping}, which are relevant to our work.

\subsection{Doping Pfaffian state with nonminimally charged anyons}

The Pfaffian state is effectively described by \cite{shi2025doping}
\begin{align}
    \mathcal{L}_\mathrm{Pf} &= - \frac{2}{4 \pi} \Tr\left[a d a + \frac{2}{3} a^3\right] + \frac{3}{4 \pi} (\Tr a) d (\Tr a) \nonumber \\ 
    &\quad - \frac{1}{2 \pi} A d (\Tr a) + \mathrm{CS}[A, g],
\end{align}
where $a$ is a dynamical $\mathrm{U}(2)$ gauge field and $A$ is the background electromagnetic field. We consider doping the semion $\alpha_{\frac{1}{2}}$ with quantum dimension $d_{\alpha_{\frac{1}{2}}} = 1$, topological spin $s_{\alpha_{\frac{1}{2}}} = \frac{1}{4}$, and charge $q_{\alpha_{\frac{1}{2}}} = \frac{e}{2}$. To incorporate the doped anyon, we add a matter term to the Lagrangian as \cite{shi2025doping}
\begin{align}
    \mathcal{L}_\mathrm{Pf} &= - \frac{2}{4 \pi} \Tr\left[a d a + \frac{2}{3} a^3\right] + \frac{3}{4 \pi} (\Tr a) d (\Tr a) \nonumber \\ 
    &\quad - \frac{1}{2 \pi} A d (\Tr a) + \mathrm{CS}[A, g] + L[\Psi, A - \Tr a]
\end{align}
where $\Psi$ is a fermionic field. The equation of motion for $\Tr a$ implies that $\Psi$ is at an effective Landau level of filling $-2$. Putting $\Psi$ in the integer quantum Hall state with filling $\nu = -2$, the Lagrangian becomes \cite{shi2025doping}
\begin{align}
    \mathcal{L}_\mathrm{SC} &= - \frac{2}{4 \pi} \Tr\left[a d a + \frac{2}{3} a^3\right] + \frac{1}{4 \pi} (\Tr a) d (\Tr a) \nonumber \\
    &\quad - \frac{1}{2 \pi} A d (\Tr a) - \mathrm{CS}[A, g].
\end{align}
This describes a charge-$2e$ superconductor with chiral central charge $c = \frac{1}{2}$.

Next, consider doping the anti-semion $\bar{\alpha}_{\frac{1}{2}}$ with quantum dimension $d_{\bar{\alpha}_{\frac{1}{2}}}$, topological spin $s_{\bar{\alpha}_{\frac{1}{2}}}$, and charge $q_{\bar{\alpha}_{\frac{1}{2}}}$. The matter term is added as \cite{shi2025doping}
\begin{align}
    \mathcal{L}_\mathrm{Pf} &= - \frac{2}{4 \pi} \Tr\left[a d a + \frac{2}{3} a^3\right] + \frac{3}{4 \pi} (\Tr a) d (\Tr a) \nonumber \\ 
    &\quad - \frac{1}{2 \pi} A d (\Tr a) + \mathrm{CS}[A, g] + L[\Phi, \Tr a]
\end{align}
where $\Phi$ is a scalar field. The equation of motion implies that $\Phi$ is at Landau level filling $\nu = -2$. Putting $\Phi$ in the bosonic integer quantum Hall state at filling $\nu = -2$, the Lagrangian becomes \cite{shi2025doping}
\begin{align}
    \mathcal{L}_\mathrm{SC} &= - \frac{2}{4 \pi} \Tr\left[a d a + \frac{2}{3} a^3\right] + \frac{1}{4 \pi} (\Tr a) d (\Tr a) \nonumber \\
    &\quad + \frac{1}{2 \pi} A d (\Tr a) + \mathrm{CS}[A, g].
\end{align}
This describes a charge-$2e$ superconductor with $c = \frac{5}{2}$.

\subsection{Doping \texorpdfstring{$\mathbb{Z}_k$}{} Read-Rezayi state}

The $\mathbb{Z}_k$ Read-Rezayi state is described by \cite{shi2025doping}
\begin{align}
    \mathcal{L}_{\mathrm{RR}_k} &= - \frac{k}{4 \pi} \Tr\left[a d a + \frac{2}{3} a^3\right] + \frac{k + 1}{4 \pi} (\Tr a) d (\Tr a) \nonumber \\
    &\quad - \frac{1}{2 \pi} A d (\Tr a) + \mathrm{CS}[A, g].
\end{align}
The anyons of the theory are labeled by $(j, n)$ with $j = 0, \frac{1}{2}, 1, \cdots, \frac{k}{2}$ and $n = 0, 1, \cdots, 2 (k + 2) - 1$ such that $j + \frac{n}{2} \in \mathbb{Z}$. For $k > 2$, there are multiple minimally charged anyons in the theory. Here, we consider doping $a_\frac{1}{k + 2}$ which is labeled by $(\frac{1}{2}, 1)$. 

Let us first assume that the paramagnetic fusion channel $(\frac{1}{2}, 1) \otimes (\frac{1}{2}, 1) \to (0, 2)$ is energetically favored. To describe the doped anyons, the matter term is introduced as \cite{shi2025doping}
\begin{align}
    \mathcal{L}_{\mathrm{RR}_k} &= - \frac{k}{4 \pi} \Tr\left[a d a + \frac{2}{3} a^3\right] + \frac{k + 1}{4 \pi} (\Tr a) d (\Tr a) \nonumber \\
    &\quad - \frac{1}{2 \pi} A d (\Tr a) + \mathrm{CS}[A, g] + L[\Phi, a].
\end{align}
The equations of motion for $a$ implies that $\Phi$ is at the effective Landau level filling $\nu = -2 (k + 2)$. By putting $\Phi$ in the bosonic integer quantum Hall state with filling $\nu = -2 (k + 2)$, the Lagrangian becomes \cite{shi2025doping}
\begin{align}
    \mathcal{L}_\mathrm{SC} &= \frac{2}{4 \pi} \Tr\left[a d a + \frac{2}{3} a^3\right] - \frac{1}{4 \pi} (\Tr a) d (\Tr a) \nonumber \\
    &\quad + \frac{1}{2 \pi} A d (\Tr a) + \mathrm{CS}[A, g].
\end{align}
This describes a charge-$2e$ superconductor with $c = - \frac{1}{2}$.

\subsection{Doping bosonic Laughlin state at filling \texorpdfstring{$\nu = \frac{1}{2}$}{}}

The bosonic Laughlin state at filling fraction $\nu = \frac{1}{2}$, doped by the semion $s_{\frac{1}{2}}$ with quantum dimension $d_{s_{\frac{1}{2}}} = 1$, topological spin $s_{\frac{1}{2}} = \frac{1}{4}$, and charge $q_{s_{\frac{1}{2}}} = \frac{1}{2}$, is described by \cite{wen2004quantum,shi2025doping}
\begin{align}
    \mathcal{L}_{\nu = \frac{1}{2}} &= - \frac{2}{4 \pi} a d a + \frac{1}{2 \pi} A d a + L[\phi, a],
\end{align}
where $a$ is a dynamical $\mathrm{U}(1)$ gauge field and $\phi$ is a scalar field. The equation of motion for $a$ implies that $\phi$ is at filling fraction $2$. Putting it in the bosonic integer quantum Hall state, we get \cite{shi2025doping}
\begin{align}
    \mathcal{L} = \frac{1}{2 \pi} A d a,
\end{align}
which describes a nonchiral charge-$e$ superconductor.

\section{Anyonic data of some topological orders} \label{app:data}

In this section, we list the anyonic data of the bosonic $\mathbb{Z}_k$ Read-Rezayi states and the $\mathbb{Z}_{2k}$ gauge theory, which appeared in the main text.

\subsection{\texorpdfstring{$\mathbb{Z}_k$}{} bosonic Read-Rezayi state}

The anyons of the $\mathbb{Z}_k$ bosonic Read-Rezayi state are described by the $\mathrm{SU}(2)_k$ modular tensor category and labeled by $j = 0, \frac{1}{2}, \cdots, \frac{k}{2}$ \cite{read1999beyond}. The topological spins of the anyons are given by \cite{seiberg2016gapped,shi2025doping}
\begin{align}
    s_j = \frac{j (j + 1)}{k + 2}.
\end{align}
The anyons obey the standard fusion rule of $\mathrm{SU}(2)$ representation with upper bound, i.e., 
\begin{align}
    j_1 \otimes j_2 \equiv \bigoplus_j N^{j_1, j_2}_j j = \bigoplus_{j = |j_1 - j_2|}^{\min(j_1 + j_2, k - j_1 - j_2)} j.
\end{align}
The quantum dimension of anyon $j_1$ is given by the largest eigenvalue of the matrix $N_{j_1}$ whose elements are given by $(N_{j_1})_{j_2, j} = N^{j_1, j_2}_j$ \cite{wen2016theory}. The $S$ matrix can be constructed by \cite{wen2016theory}
\begin{align}
    S_{j_1, j_2} = \frac{1}{D} \sum_j N^{j_1, j_2}_j e^{2 \pi i (s_{j_1} + s_{j_2} - s_j)} d_j,
\end{align}
where $D = \sqrt{\sum_j d_j^2}$ is the total quantum dimension. From the Lagrangian Eq.~\eqref{eq:bRR_Lagrangian}, one can find that the $j = \frac{1}{2}$ anyon considered in the main text has charge $\frac{e}{2}$.

\subsection{\texorpdfstring{$\mathbb{Z}_{2k}$}{} gauge theory}

The $\mathbb{Z}_{2k}$ gauge theory $D(\mathbb{Z}_{2k})$ has $4 k^2$ anyon types, which are labeled by $(p, q)$ for $p, q = 0, 1, \cdots, 2k - 1$. The anyons are all Abelian, and their topological spins are given by \cite{lin2014generalizations}
\begin{align}
    s_{(p, q)} = e^{\pi i \frac{pq}{k}}.
\end{align}
In terms of the $K$-matrix formulation \cite{wen1992classification}, the charge-neutral $D(\mathbb{Z}_{2k})$ is characterized by \cite{seiberg2016gapped}
\begin{align}
    K = 
    \begin{pmatrix}
        0 & 2k \\
        2k & 0
    \end{pmatrix}, \quad 
    \mathbf{t} = 
    \begin{pmatrix}
        0 \\ 0
    \end{pmatrix}.
\end{align}
Anyons are labeled by the integer vectors $\mathbf{l} = (l_1, l_2)^\mathsf{T}$ such that $0 \leq l_1, l_2 \leq 2k - 1$. The minimal ``electric'' excitation $e$ appeared in Eq.~\eqref{eq:condensable_bosonic_RR_SC} corresponds to $\mathbf{l}_e = (1, 0)^\mathsf{T}$. It is straightforward to check that $e$ is indeed bosonic: $\pi \mathbf{l}_e^\mathsf{T} K^{-1} \mathbf{l}_e = 0$.

\bibliography{ref}

\end{document}